\documentclass[letter]{aa}

\usepackage{graphicx}
\usepackage{txfonts}
\usepackage{ulem}
\usepackage{makecell}
\usepackage{booktabs}
\usepackage{chngcntr}
\counterwithin{table}{section}

\begin{document}

   \title{Evidence of a non-equipartition energy regime in 1803+784}

   \subtitle{Core-shift and Faraday rotation measurements from simultaneous multi-frequency polarimetric VGOS observations}

   \author{V. Pérez-Díez\inst{1,4},
    I. Martí-Vidal\inst{2,3},
    E. Albentosa-Ruiz\inst{2},
    R. Bachiller\inst{1}}

   \institute{
   Observatorio Astron\'omico Nacional (OAN-IGN), Calle Alfonso XII 3, 28014 Madrid, Spain
   \and Dpt. Astronomia i Astrof\'isica, Universitat de Val\`encia, C/ Dr. Moliner 50, 46120 Burjassot, Spain
   \and Observatori Astron\`omic, Universitat de Val\`encia, C/ Cat. Jos\'e Beltr\'an 2, 46980 Paterna, Spain
   \and Centro de Desarrollos Tecnológicos, Observatorio de Yebes (IGN), 19141 Yebes, Guadalajara, Spain
   }

   \date{Received 28 May, 2025; accepted 30 July, 2025}

\abstract
{Compact jets from active galactic nuclei (AGN) are commonly assumed to be in equipartition between particle and magnetic-field energy densities at the regions where the radio emission dominates at centimetre wavelengths. This assumption has significant implications for both jet physics and the accuracy of VLBI-based astrometry and geodesy.}
{We tested the validity of the energy equipartition hypothesis in AGN cores at centimetre wavelengths by analysing the blazar 1803+784 using simultaneous broadband full-polarization observations with the VLBI Global Observing System (VGOS).}
{We present VGOS observations of the blazar 1803+784 covering the 3-11\,GHz frequency range. The data were processed using a dedicated calibration pipeline, followed by model fitting and multi-frequency imaging analysis. We measured the frequency-dependent core shift and mapped the spectral index and rotation measure (RM) across the source.}
{We find a core-shift power-law index of $k_r = 0.73^{+0.12}_{-0.19}$, significantly deviating from the expected equipartition value of $k_r = 1$. This indicates that either the equipartition condition or the conical jet geometry, or both, are not fulfilled in the centimetre-wavelength core region. The wide frequency coverage of VGOS also allows us to decouple the Faraday rotation of the core into an internal component ($RM_I = 121 \pm 8$ rad m$^{-2}$, produced in the core region) and an external component ($RM_E = -44 \pm 9$ rad m$^{-2}$, associated with a distant, extended medium that may also affect the polarization in downstream regions of the jet at larger scales).}
{These results demonstrate the power of VGOS for high-fidelity simultaneous multi-frequency polarimetric studies of compact AGN jets, and underline the need to account for non-equipartition effects in both jet astrophysics and geodetic VLBI.}


   \maketitle

\section{Introduction}

Compact jets from active galactic nuclei (AGN), particularly those associated with blazars and radio-loud quasars, are widely used in geodetic very long baseline interferometry (VLBI). However, the physical conditions and structure of these jets remain under active study. Jet models typically assume equipartition between the energy densities of relativistic particles and magnetic fields \citep[e.g.][]{Plavin2019, OSullivan, OSullivan2009}. This assumption has been tested in some sources and found to be consistent with observations, as shown in \citet{Hada2011} and \citet{Sokolovsky2011}. The equipartition hypothesis implies a specific frequency dependence for the apparent core position, characterized by a power-law shift with exponent $k_r \approx 1$ \citep{Blandford,Lobanov1998}:

\begin{equation}
r_{\text{core}} \propto \nu^{-1/k_r}
    \label{eq:CoreShift}
\end{equation}

When $k_r = 1$, the frequency-dependent shift in the core position does not introduce a group delay. Therefore, group-delay VLBI astrometry is not affected by frequency-dependent core shifts \citep{Porcas2009}. However, when $k_r \neq 1$, the group-delay position becomes frequency-dependent and deviates from the physical jet origin. This can introduce systematic errors in astrometric VLBI solutions. \citet{Kovalev2008} showed that such core shifts can bias the alignment between radio and optical reference frames, compromising the precision of celestial reference frame realizations such as the International Celestial Reference Frame \citep[ICRF;][]{Ma1998, Charlot2020} and the \textit{Gaia} catalogue \citep{Gaia2016, Lindegren2018, Gaia2023}.

The advent of the VLBI Global Observing System (VGOS), the next generation of geodetic VLBI observations, developed under the coordination of the International VLBI Service for Geodesy and Astrometry (IVS)\footnote{\url{https://ivscc.gsfc.nasa.gov/index.html}} \citep{Nothnagel2017}, allows a robust test of the equipartition hypothesis on a sample of radio-loud AGN routinely used for geodetic studies. Designed to achieve station position accuracies at the millimetre level, and with the capability for simultaneous multi-frequency observations, VGOS employs ultra-wideband receivers capable of covering approximately 2-15\,GHz. The broad frequency coverage enables the measurement of group delays with picosecond-level precision, a critical requirement for high-accuracy geodesy and astrometry \citep{VGOSRef, Niell2018}.

Verifying whether AGN cores adhere to the expected $k_r = 1$ regime is not only important for understanding jet physics, but is also crucial for ensuring the accuracy of geodetic measurements \citep[e.g.][]{Porcas2009, Ming2022} and for aligning radio and optical reference frames \citep{Kovalev2008, Ming2021b}.

In this Letter we present an analysis of high-sensitivity VGOS observations of the prototypical blazar 1803+784. These multi-frequency data allowed us to derive a precise measurement of the core shift as a function of frequency. Additionally, we exploited the broad frequency coverage of VGOS to measure both internal and external Faraday rotation in the jet of 1803+784. 

Finally, we produced high-resolution multi-frequency images of this compact jet, and created detailed maps of the spectral index and rotation measure. Furthermore, accurately characterizing the source structure in each band and its alignment through the core shift makes it possible to correct the structure effects described by \citet{Ming2021a}, as demonstrated by the direct application of this pipeline to VGOS data in \citet{Jaron2025}.

\section{Observations}
  
For this work we used observations from the IVS experiment VO2187, observed on 6-7 July 2022, involving eight antennas: Goddard (GS), Ishioka (IS), Kokee (K2), McDonald (MG), Onsala East (OE), Onsala West (OW), Westford (WF), and Yebes (YJ). Full details of the frequency set-up and observation strategy can be found in \cite{PerezDiez2024}.

The data is divided into four frequency bands labelled A, B, C, and D, following the naming convention adopted by the VGOS community. Each band consists of eight spectral windows (spws) of 32\,MHz and spans 0.5\,GHz. The bands are centred at 3.25\,GHz (A), 5.5\,GHz (B), 6.75\,GHz (C), and 10.5\,GHz (D). In this Letter we focus on the source that was most extensively observed, the prototypical blazar 1803+784.

\section{Methods}

\subsection{Calibration}
\label{sec:calibration}

The calibration strategy used in this work is described in \citet{PerezDiez2024}. The first step consisted in converting the linear-polarization visibilities into circular basis using \texttt{PolConvert} \citep{MartiVidal2016}. After this, we applied a wide-band global fringe-fitting algorithm that solves simultaneously for dispersive and non-dispersive delays across the full VGOS frequency range. This method is optimized for the VGOS multi-band design and uses ionospheric models from GNSS IONEX maps produced by Jet Propulsion Laboratory (JPL)\footnote{https://cddis.nasa.gov/archive/gnss/products/ionex} \citep{Martire2024} to improve convergence.

The amplitude calibration was refined through a hybrid mapping procedure using \texttt{Difmap} \citep{Difmap}, combining iterative CLEAN deconvolution and self-calibration. We first performed several iterations of phase self-calibration, with decreasing solution intervals (ranging from 30 minutes to 1 minute), using total intensity (Stokes I) CLEAN models. Then, a few extra iterations of amplitude and phase self-calibration were performed, also based on Stokes I, using longer solution intervals (30$-$20 minutes). The instrumental polarization (D-terms) was refined using the \texttt{PolSolve} software \citep{polsolve}, which performs a global fit of the radio interferometer measurement equation (RIME), formulated by \citet{Hamaker1996}, to minimize instrumental polarization leakage. This step ensures reliable full-Stokes calibration across the entire VGOS bandwidth, correcting for residual leakage effects that may remain after the initial polarization conversion.

\subsection{Model fitting}
\label{sec:modelfit}
To characterize the source structure, we performed model fitting to the visibilities following a procedure based on \cite{ref:uvmultifit}. The source brightness model is a combination of several Gaussian components: a central elliptical Gaussian for the optically thick core, and two Gaussians to represent the extended optically thin jet emission. The fitting was initially performed in total intensity (Stokes I) at C band (6.75\,GHz), which provides the best compromise between resolution and signal-to-noise ratio. 

To model the frequency-dependent core shift, the positions and sizes (but not the flux densities) of every optically thin jet component were kept fixed across bands and the position of the core component was left free. In this fit, the Gaussian shape parameters of the core were also held fixed to minimize systematics related to jet blending at the lower frequencies.

Lastly, the polarization structure (Stokes Q, U, and V) was modelled using the same Gaussian components fitted to Stokes I. In this case, only the flux densities of the individual components were allowed to vary to  ensure consistency between the total intensity and polarimetric models across all frequencies.

\subsection{Multi-frequency imaging}
\label{sec:Imaging}

To reconstruct the source structure across the wide frequency coverage provided by VGOS, we employed a regularized maximum likelihood (RML) approach using the \texttt{eht-imaging} software \citep{Chael2016, Chael2018}. This technique uses closure phases, log closure amplitudes, and visibility amplitudes as observables. This combination  is robust against station-based calibration errors, and thus simplifies the imaging process.

The imaging was carried out in multi-frequency mode, recently implemented in \texttt{ehtim} \citep{Chael2023}. This algorithm performs a log-log Taylor expansion of the image intensity as a function of frequency around a reference frequency $\nu_0$, which enables the simultaneous reconstruction of   a total intensity map and a spectral index map, thus producing aligned images across frequencies. Maps for the other Stokes parameters were deconvolved by imaging each band separately with \texttt{ehtim}.

\section{Results}

\subsection{Core shift and non-equipartition regime}

\begin{figure}[!ht]
   \centering
   \includegraphics[width=\hsize]{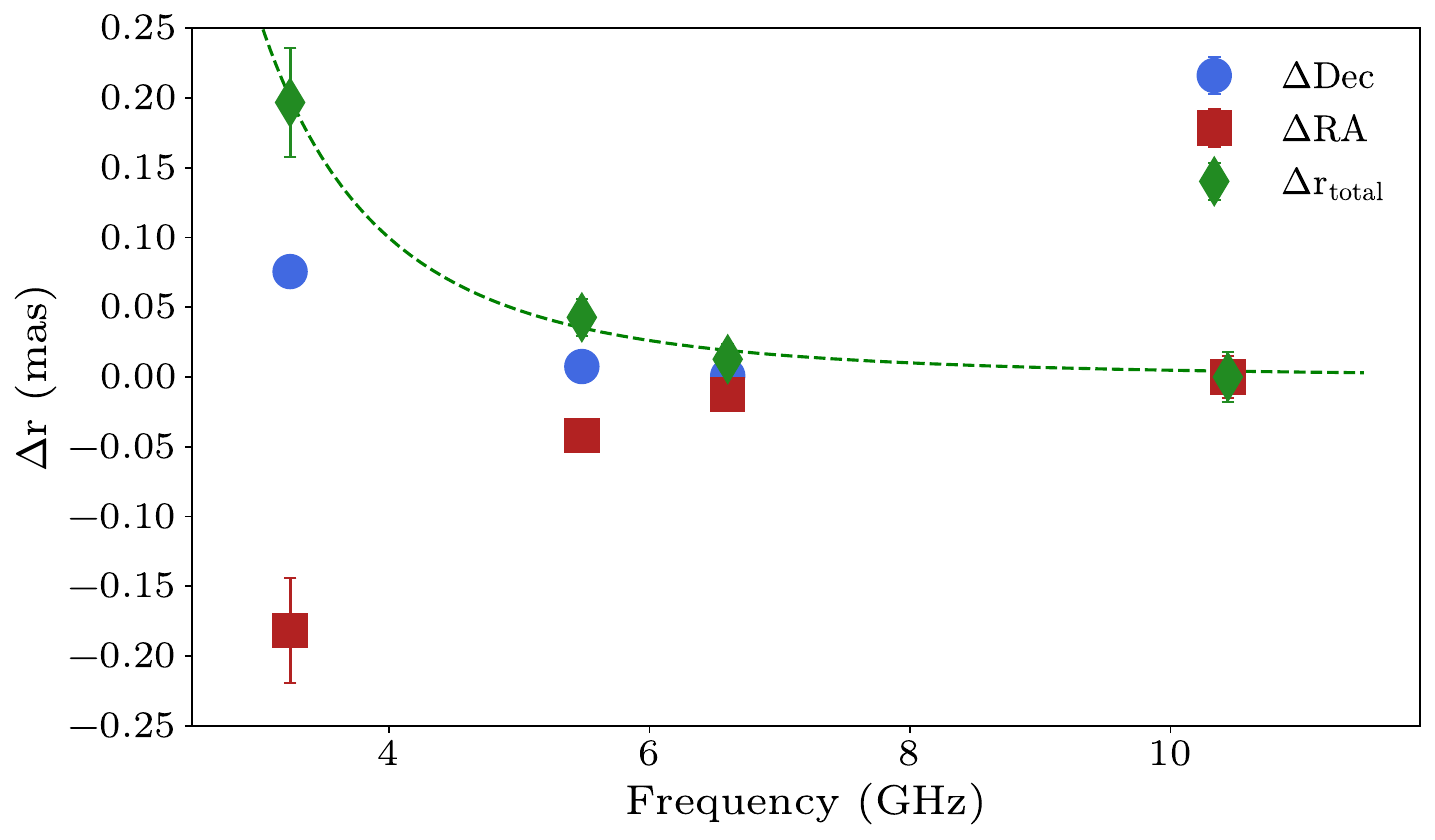}
   \caption{Measured relative core shift as a function of frequency for 1803+784. The total shift vector, along with its components in right ascension and declination, are shown. The error bars represent 3$\sigma$ uncertainties. The best-fit model is shown as a dashed green line.}
   \label{fig:core_shift}
\end{figure}

We have measured a core shift of approximately 0.20 milliarcseconds (mas) between 3 and 11\,GHz. In Fig.~\ref{fig:core_shift} we show the core positions relative to the one at the highest frequency (10.66\,GHz). The shift is decomposed into its right ascension (RA) and declination (Dec) components, both exhibiting consistent trends with frequency. The total core-shift vector, $\Delta r$, follows a clear power-law behaviour with frequency. The wide bandwidth of VGOS, with eight  spws per band, allows for relatively small uncertainties in the measurements. The uncertainty in band A is larger due to greater scatter among spws, possibly caused by its bigger beam.

By fitting the frequency dependence of the core shift described in Eq.~\ref{eq:CoreShift}, we obtain a best-fit power-law index of $k_r = 0.73^{+0.12}_{-0.19}$. This result represents a significant deviation from the standard equipartition condition ($k_r \sim 1$), which assumes a balance between the energy densities of relativistic particles and magnetic fields in a conical, freely expanding jet \citep{Lobanov1998}. A value of $k_r < 1$ indicates that the equipartition condition, the conical jet geometry, or both may not hold at the region we are observing at these frequencies. These findings underline the necessity of carefully evaluating the core-shift behaviour in blazars when deriving physical jet parameters, such as magnetic field strength, and when conducting geodetic studies.

\subsection{Rotation measure}

The wide simultaneous frequency coverage provided by VGOS allows us, for the first time, to separate the internal and external contributions of the Faraday rotation observed in 1803+784. While a distant plasma screen introduces an external rotation measure, $RM_E$, that rotates the “electric vector position angle (EVPA), $\chi$, as $\Delta \chi = RM_E \, \Delta \lambda^2$\,, an internal (and/or surrounding) screen around the optically thick core produces a rotation measure $RM_I$ that depends on frequency as $RM_I \propto \nu^{a}$, as derived in \citet{Jorstad2007}. The exponent $a$ describes the radial dependence of the electron density, assuming a power-law decrease with distance from the black hole: $n_e \propto r^{-a}$ \citep{Hovatta2019}. Reported values of the exponent $a$ in the literature typically range from 0.4 to 4, with average values of around 2 \citep{Jorstad2007, OSullivan2009, Kravchenko2017, Hovatta2019}. This average value of $a=2$ suggests that the Faraday rotation takes place in a sheath surrounding a conically expanding jet. Lower values could be attributed to a more collimated jet structure \citep{Hovatta2019}, different geometry of the jet, flaring activity, or a possible filamentary structure of the Faraday screen \citep{Kravchenko2017}.

\begin{figure}[!ht]
   \centering
   \includegraphics[width=0.85\hsize, trim=0.0cm 0.3cm 0.0cm 0]{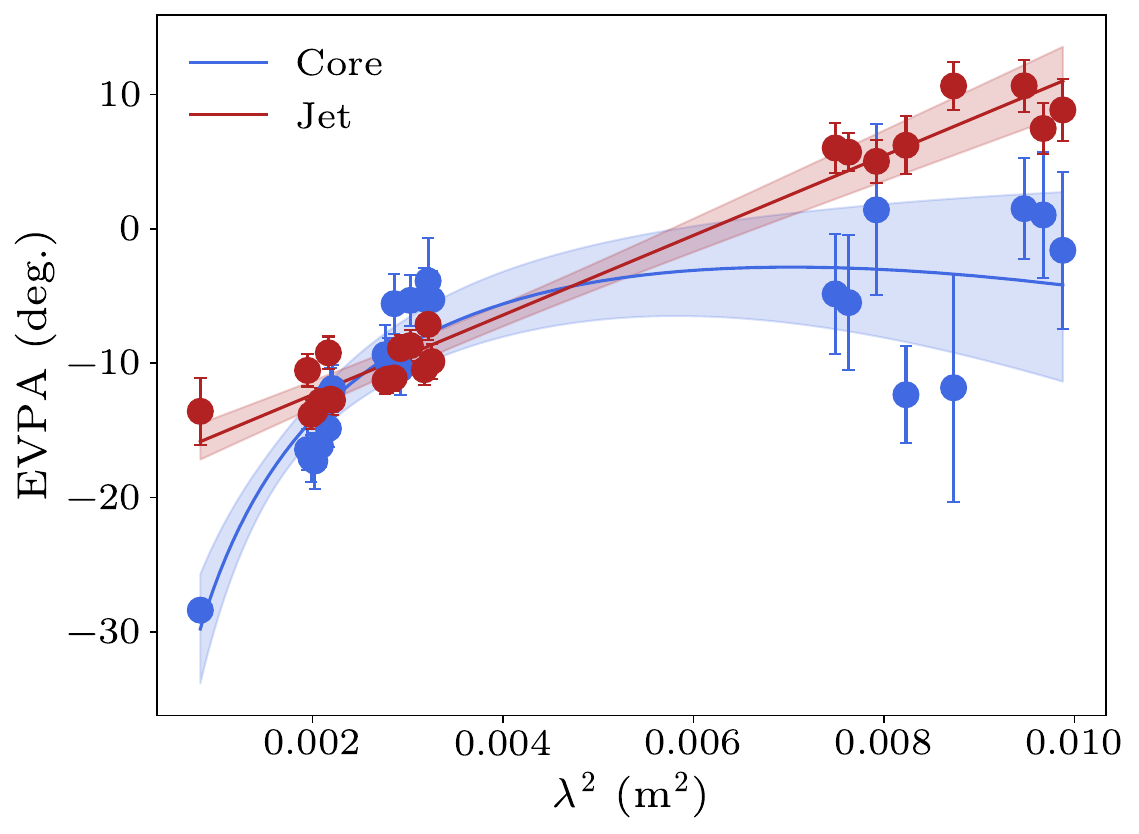}
   \caption{Fit of the polarization angle as a function of $\lambda^2$ for the core and jet regions of 1803+784. The shaded regions represent the 3$\sigma$ confidence intervals of the model fits. Band D data have been averaged because of the low S/N in this band.}
   \label{fig:RM_fit}
\end{figure}

By fitting the polarization angle, $\chi$, of the VLBI core as a function of $RM_E$ and $RM_I$ (see Fig.~\ref{fig:RM_fit}), we measure an external Faraday rotation of $RM^{\rm core}_E = -44 \pm 9$ rad m$^{-2}$ and an internal rotation of $RM^{\rm core}_I = 121 \pm 8$ rad m$^{-2}$ at the reference frequency of 5.5\,GHz.  When fitting for the exponent of $RM^{\rm core}_I$ with $\nu$, the result is $a = 2.18 \pm 0.95$. Due to large uncertainties, we cannot confirm this value, but it seems to be consistent with the theoretical expectation for a common expanding jet.

On the other hand, the plasma in the jet components is optically thin, so there is no core shift and the effects of $RM^{\rm jet}_E$ and $RM^{\rm jet}_I$ both follow a $\lambda^2$ relation. In this case, the fitted rotation measure is $RM^{\rm jet}_{\rm tot} = RM^{\rm jet}_E + RM^{\rm jet}_I = 49 \pm 2$ rad m$^{-2}$, and $\chi$ follows a clean $\lambda^2$ dependence, consistent with the presence of a total optically thin Faraday screen (Fig.~\ref{fig:RM_fit}). 

If we consider that the effect of the distant plasma screen does not change between the core and the inner jet, we can further infer the $RM^{\rm jet}_I$ contribution by taking $RM^{\rm jet}_E \sim RM^{\rm core}_E$. With this assumption, the internal rotation measure in the optically thin jet is approximately $RM^{\rm jet}_I \sim 93 \pm 10$ rad m$^{-2}$, which is lower than $RM^{\rm core}_I$. This is in agreement with a jet magnetic field that decreases with distance to the black hole, as is implied in the model, $RM \propto \nu^p$, used to fit the $\chi$ at the core. Hence, the Faraday rotation at the core and the jet are both self-consistent with the picture of having two Faraday screens, one distant and extended and a second one located in the immediate jet surroundings (and/or internal to the jet itself).

\subsection{Multi-frequency image}

\begin{figure}[!t]
\centering
\includegraphics[width=1.01\hsize, trim=0.3cm 0.3cm 0.3cm 0, clip]{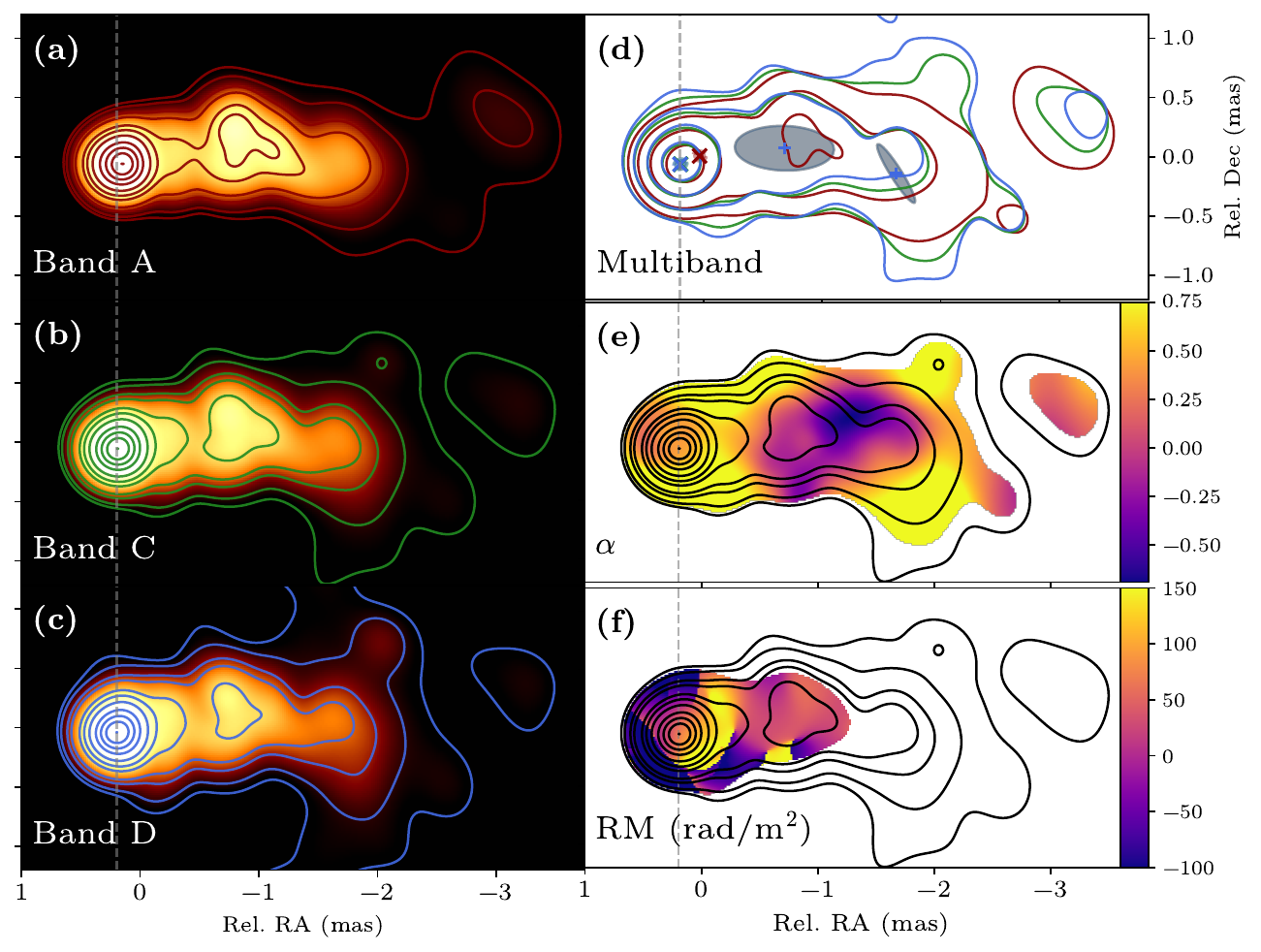}
   \caption{Multi-frequency polarimetric image of 1803+784. Panels (a)--(c) show the total intensity maps at bands A, C, and D in logarithmic scale, normalized to their peak. The dashed vertical line indicates the core position at band D, marking the core shift. Panel (e) displays all contours from the four bands superposed, along with the core and jet Gaussian models. Panel (f) shows the spectral index map ($\alpha$). Panel (g) presents the rotation measure (RM) map. Only regions above 0.8\% of the peak intensity are included.}
   \label{fig:all_panels}
\end{figure}

Figure~\ref{fig:all_panels} presents the multi-frequency imaging results for 1803+784. 
Panels (a)\textendash(c) show the total intensity (Stokes I) maps at three frequency bands (A, C, and D), with contour lines superimposed over colour raster images. The contours are drawn at 0.08, 0.2, 0.5, 1, 5, 10, 20, 40, 60, and 80\% of the peak intensity in each band. The colour raster highlights the detailed jet structure at each frequency.

A vertical dashed grey line marks the RA position of the core in band D. A shift of the core to the west with increasing frequency is clearly noticeable and compatible with the one obtained in the model fitting.  

Panel (e) overlays the total intensity contours from all bands, allowing a direct visualization of the core shift. The Gaussian components derived from the model fitting described in Sect.~\ref{sec:modelfit} are also plotted, with colours corresponding to each band: red (A), orange (B), green (C), and purple (D).

Panel (f) displays the spectral index ($\alpha$) map, computed from the multi-frequency imaging described in Sect.~\ref{sec:Imaging}. The core region exhibits an inverted spectrum ($\alpha > 0$) consistent with synchrotron self-absorption, while the jet shows a steep spectrum ($\alpha < 0$), as expected for optically thin synchrotron emission.

Finally, panel (g) presents the rotation measure (RM) map derived from the multi-band polarimetric data. The RM structure reveals a complex distribution, with higher RM values near the core and lower values along the jet.

\section{Discussion and conclusions}

Our results provide the first direct evidence for a significant departure from equipartition conditions in the compact jet of 1803+784, based on simultaneous multi-frequency and full-polarization VLBI observations. The measured core-shift power-law index ($k_r = 0.73^{+0.12}_{-0.19}$) strongly deviates from the standard equipartition prediction ($k_r \sim 1$; \citealt{Lobanov1998}), indicating that either the energy density balance between particles and magnetic fields, the conical jet geometry, or both assumptions, are not fully satisfied in the innermost jet regions during this epoch.

\citet{Kovalev2008} obtained a core-shift value of 0.72\,mas between 2.3 and 8.6\,GHz for 1803+784. Extrapolating our fit to a frequency of 2.3\,GHz, a core-shift value of 0.61\,mas is obtained, although this difference may be due to the fact that the observations are separated by 17 years.

More recent studies of another source, NGC 315, obtain $k_r$ values similar to those obtained in this work. \citet{Boccardi2021} fits core-shift values to six different frequencies (in a range similar to ours), obtaining a value of $k_r = 0.84 \pm 0.06$. However, a reanalysis of the data by \citet{Ricci2022}, where they add an additional point at 43\,GHz, fits two $k_r$ values for different frequency ranges: $k_r = 0.93 \pm 0.01$ for frequencies lower than 10\,GHz and $k_r = 0.57 \pm 0.17$ for higher frequencies. According to their interpretation, these two distinct regimes may test the parabolic and conical zones of the jet, that is, the transition between a non-equipartition zone and an equipartition zone.

Following this hypothesis, we   performed the same 10\,GHz separation. Fitting without data at D band (10.5\,GHz) we obtain a value of $k_r = 0.85^{+0.03}_{-0.06}$ for the three frequencies lower than 10\,GHz. This value is closer to unity, without reaching equipartition, suggesting that the scenario proposed by \citet{Ricci2022} may also apply to 1803+784, but the transition would occur at a lower frequency. Observations at frequencies below our band A would be necessary to confirm this. We also note that the observations of \citet{Boccardi2021, Ricci2022} are separated by more than 20 years, while the VGOS observations presented here are truly simultaneous.

The ability to separate internal and external Faraday rotation contributions, thanks to VGOS's wide instantaneous bandwidth, represents another major advance. Our measurement of a positive internal RM and a negative external RM near the core supports the presence of two distinct Faraday screens, likely associated with a stratified plasma layer surrounding the base of the jet. Our inferred external Faraday rotation ($RM_E = -44 \pm 9$ rad m$^{-2}$ after correcting for internal contributions) is broadly consistent with previous measurements at larger scales  (i.e. farther downstream from the core) by \citet{Gabuzda2003}.

The spectral index and RM maps derived from our multi-band imaging show an excellent agreement with previous studies such as the MOJAVE programme \citep{Hovatta2012} This confirms the reliability of VGOS observations and our calibration pipeline in obtaining astrophysical results.

These results highlight the potential of VGOS for high-fidelity multi-frequency full-polarization studies of compact AGN jets. In particular, they demonstrate the importance of accounting for non-equipartition effects when deriving jet parameters or performing high-precision VLBI astrometry.

 \begin{acknowledgements}
This work has been partially supported by the Generalitat Valenciana GenT Project CIDEGENT/2018/021 and by the MICINN Research Projects PID2019-108995GB-C22 and PID2022-140888NB-C22.

This work has been supported by the grant PRE2020-092200 funded by MCIN/AEI/ 10.13039/501100011033 and by ESF invest in your future.

\end{acknowledgements}

\bibliographystyle{aa}
\bibliography{vo2187}

\begin{appendix} 

\section{Core-shift fitting}
\label{sec:CoreShiftFit}

To determine the power-law index $k_r$ describing the frequency dependence of the core shift, we fitted the averaged total shift values $\Delta r$ as a function of frequency using the model $\Delta r(\nu) = A \cdot \nu^{-1/k_r}$. The fit was performed using a weighted least-squares minimization.

To estimate the uncertainty of $k_r$, we generated Monte Carlo simulations of the fit parameters, sampling from the multi-variate normal distribution defined by the best-fit covariance matrix. The asymmetric uncertainties were derived from the 16th and 84th percentiles of the resulting distribution.

The data used for the fit is presented in Table~\ref{tab:core_shift}. For each band, the spw measurements were averaged, and the fitting was carried out using these averaged values, which are emphasized in boldface.

\begin{table}[h]
\centering
\caption{Core-shift measurements with associated uncertainties. }
\label{tab:core_shift}
\setlength{\tabcolsep}{4pt}
\begin{tabular}{cccc}
\toprule
\makecell{Freq\\(GHz)} & \makecell{RA\\(mas)} & \makecell{Dec\\(mas)} & \makecell{Total\\(mas)} \\
\midrule
3.016 & $-0.00 \pm 0.03$ & $0.090 \pm 0.008$ & $0.09 \pm 0.03$ \\
3.048 & $-0.29 \pm 0.04$ & $0.068 \pm 0.008$ & $0.29 \pm 0.04$ \\
3.080 & $-0.04 \pm 0.03$ & $0.105 \pm 0.008$ & $0.11 \pm 0.03$ \\
3.208 & $-0.06 \pm 0.02$ & $0.091 \pm 0.007$ & $0.11 \pm 0.02$ \\
3.304 & $-0.239 \pm 0.018$ & $0.060 \pm 0.005$ & $0.25 \pm 0.02$ \\
3.368 & $-0.159 \pm 0.017$ & $0.078 \pm 0.005$ & $0.18 \pm 0.018$ \\
3.432 & $-0.224 \pm 0.018$ & $0.079 \pm 0.005$ & $0.24 \pm 0.019$ \\
3.464 & $-0.280 \pm 0.018$ & $0.063 \pm 0.005$ & $0.29 \pm 0.019$ \\
\textbf{3.240} & {\boldmath$\mathbf{-0.18} \pm \mathbf{0.03}$} & {\boldmath$\mathbf{0.076} \pm \mathbf{0.011}$} & {\boldmath$\mathbf{0.20} \pm \mathbf{0.04}$} \\
5.256 & $-0.034 \pm 0.008$ & $0.005 \pm 0.005$ & $0.034 \pm 0.010$ \\
5.288 & $-0.052 \pm 0.008$ & $0.012 \pm 0.004$ & $0.054 \pm 0.010$ \\
5.320 & $-0.042 \pm 0.008$ & $0.011 \pm 0.004$ & $0.044 \pm 0.009$ \\
5.448 & $-0.035 \pm 0.006$ & $0.013 \pm 0.003$ & $0.038 \pm 0.007$ \\
5.544 & $-0.045 \pm 0.006$ & $0.007 \pm 0.003$ & $0.046 \pm 0.007$ \\
5.608 & $-0.054 \pm 0.006$ & $0.004 \pm 0.003$ & $0.054 \pm 0.006$ \\
5.672 & $-0.027 \pm 0.007$ & $0.003 \pm 0.004$ & $0.027 \pm 0.008$ \\
5.704 & $-0.045 \pm 0.005$ & $0.005 \pm 0.003$ & $0.045 \pm 0.006$ \\
\textbf{5.480} & {\boldmath$\mathbf{-0.042} \pm \mathbf{0.012}$} & {\boldmath$\mathbf{0.008} \pm \mathbf{0.006}$} & {\boldmath$\mathbf{0.043} \pm \mathbf{0.013}$} \\
6.376 & $-0.019 \pm 0.005$ & $0.004 \pm 0.003$ & $0.020 \pm 0.006$ \\
6.408 & $-0.027 \pm 0.006$ & $-0.002 \pm 0.004$ & $0.027 \pm 0.007$ \\
6.440 & $-0.014 \pm 0.006$ & $-0.002 \pm 0.004$ & $0.014 \pm 0.007$ \\
6.568 & $-0.013 \pm 0.005$ & $0.001 \pm 0.003$ & $0.013 \pm 0.005$ \\
6.664 & $-0.010 \pm 0.005$ & $0.002 \pm 0.003$ & $0.010 \pm 0.006$ \\
6.728 & $-0.010 \pm 0.005$ & $0.001 \pm 0.003$ & $0.010 \pm 0.006$ \\
6.792 & $-0.006 \pm 0.006$ & $0.001 \pm 0.004$ & $0.006 \pm 0.007$ \\
\textbf{6.600} & {\boldmath$\mathbf{-0.013} \pm \mathbf{0.009}$} & {\boldmath$\mathbf{0.010} \pm \mathbf{0.006}$} & {\boldmath$\mathbf{0.013} \pm \mathbf{0.011}$} \\
10.216 & $-0.002 \pm 0.009$ & $-0.000 \pm 0.006$ & $0.002 \pm 0.010$ \\
10.248 & $-0.001 \pm 0.008$ & $-0.001 \pm 0.005$ & $0.001 \pm 0.009$ \\
10.280 & $0.001 \pm 0.008$ & $-0.001 \pm 0.005$ & $0.001 \pm 0.010$ \\
10.408 & $0.000 \pm 0.008$ & $0.001 \pm 0.005$ & $0.001 \pm 0.010$ \\
10.504 & $0.001 \pm 0.008$ & $0.001 \pm 0.005$ & $0.002 \pm 0.010$ \\
10.568 & $0.000 \pm 0.008$ & $-0.000 \pm 0.005$ & $0.000 \pm 0.010$ \\
10.632 & $0.000 \pm 0.009$ & $0.000 \pm 0.006$ & $0.000 \pm 0.010$ \\
10.664 & $-0.000 \pm 0.010$ & $0.000 \pm 0.006$ & $0.000 \pm 0.012$ \\
\textbf{10.440} & {\boldmath$\mathbf{-0.000} \pm \mathbf{0.015}$} & {\boldmath$\mathbf{-0.000} \pm \mathbf{0.010}$} & {\boldmath$\mathbf{0.000} \pm \mathbf{0.018}$} \\
\bottomrule
\end{tabular}
\tablefoot{Bold rows show the averaged data plotted.}
\end{table}

\clearpage

\section{Rotation measure data}
\label{sec:RMFit}

To estimate the Faraday rotation measures (RM) for the core and jet components, we fitted the observed electric vector position angles (EVPA) as a function of wavelength squared ($\lambda^2$) using a model that accounts for both external and internal Faraday rotation. For the core component, the fit includes an intrinsic rotation term with a power-law index describing the frequency dependence of the internal Faraday rotation.

The fits were performed by minimizing the residuals using a non-linear least-squares method. To quantify the uncertainties of the fitted parameters, we generated Monte Carlo simulations by randomly perturbing the EVPA data within their measurement uncertainties. The final uncertainties were estimated from the standard deviation of the resulting parameter distributions.

The individual EVPA measurements used for the fits are summarized in Table~\ref{tab:core_evpa_simple} (for the core) and Table~\ref{tab:jet_evpa_simple} (for the jet).

\begin{table}[h]
\centering
\caption{EVPA data for the Core component.}
\label{tab:core_evpa_simple}
\setlength{\tabcolsep}{4pt}
\begin{tabular}{cccc}
\toprule
Band & Freq (GHz) & \makecell{$\lambda^2$ (m$^2$)} & \makecell{EVPA (deg)} \\
\midrule
A & 3.0163 & 0.00988 & $-1.60 \pm 5.85$ \\
  & 3.0483 & 0.00967 & $1.03 \pm 4.68$ \\
  & 3.0803 & 0.00947 & $1.50 \pm 3.75$ \\
  & 3.2083 & 0.00873 & $-11.83 \pm 8.48$ \\
  & 3.3043 & 0.00823 & $-12.35 \pm 3.62$ \\
  & 3.3683 & 0.00792 & $1.41 \pm 6.36$ \\
  & 3.4323 & 0.00763 & $-5.49 \pm 5.00$ \\
  & 3.4643 & 0.00749 & $-4.84 \pm 4.48$ \\
B & 5.2563 & 0.00325 & $-5.27 \pm 2.13$ \\
  & 5.2883 & 0.00321 & $-3.88 \pm 3.17$ \\
  & 5.3203 & 0.00318 & $-5.22 \pm 2.30$ \\
  & 5.4483 & 0.00303 & $-5.33 \pm 1.90$ \\
  & 5.5443 & 0.00292 & $-10.44 \pm 1.91$ \\
  & 5.6083 & 0.00286 & $-5.57 \pm 2.23$ \\
  & 5.6723 & 0.00279 & $-9.94 \pm 1.83$ \\
  & 5.7043 & 0.00276 & $-9.37 \pm 2.18$ \\
C & 6.3763 & 0.00221 & $-11.89 \pm 1.74$ \\
  & 6.4083 & 0.00219 & $-12.16 \pm 1.80$ \\
  & 6.4403 & 0.00217 & $-14.87 \pm 1.40$ \\
  & 6.5683 & 0.00208 & $-16.16 \pm 1.59$ \\
  & 6.6643 & 0.00202 & $-17.28 \pm 2.06$ \\
  & 6.7283 & 0.00199 & $-17.08 \pm 1.76$ \\
  & 6.7923 & 0.00195 & $-16.43 \pm 1.52$ \\
D & 10.2163 & 0.00086 & $-26.79 \pm 1.31$ \\
  & 10.2483 & 0.00086 & $-25.95 \pm 1.21$ \\
  & 10.2803 & 0.00085 & $-28.56 \pm 1.24$ \\
  & 10.4083 & 0.00083 & $-28.39 \pm 1.19$ \\
  & 10.5043 & 0.00081 & $-29.88 \pm 1.32$ \\
  & 10.5683 & 0.00080 & $-30.05 \pm 1.29$ \\
  & 10.6323 & 0.00080 & $-27.84 \pm 1.11$ \\
  & 10.6643 & 0.00079 & $-29.64 \pm 1.40$ \\
D (avgd)  & 10.4563 & 0.00082 & $-28.39 \pm 0.72$ \\
\bottomrule
\end{tabular}
\end{table}

\begin{table}[h]
\centering
\caption{EVPA data for the Jet component.}
\label{tab:jet_evpa_simple}
\setlength{\tabcolsep}{4pt}
\begin{tabular}{cccc}
\toprule
Band & Freq (GHz) & \makecell{$\lambda^2$ (m$^2$)} & \makecell{EVPA (deg)} \\
\midrule
A & 3.0163 & 0.00988 & $8.85 \pm 2.32$ \\
 & 3.0483 & 0.00967 & $7.47 \pm 1.88$ \\
 & 3.0803 & 0.00947 & $10.65 \pm 1.94$ \\
 & 3.2083 & 0.00873 & $10.64 \pm 1.81$ \\
 & 3.3043 & 0.00823 & $6.22 \pm 2.15$ \\
 & 3.3683 & 0.00792 & $5.02 \pm 1.59$ \\
 & 3.4323 & 0.00763 & $5.70 \pm 1.42$ \\
 & 3.4643 & 0.00749 & $6.02 \pm 1.85$ \\
B & 5.2563 & 0.00325 & $-9.88 \pm 1.28$ \\
 & 5.2883 & 0.00321 & $-7.11 \pm 1.14$ \\
 & 5.3203 & 0.00318 & $-10.49 \pm 1.12$ \\
 & 5.4483 & 0.00303 & $-8.69 \pm 1.18$ \\
 & 5.5443 & 0.00292 & $-8.91 \pm 1.04$ \\
 & 5.6083 & 0.00286 & $-11.10 \pm 0.95$ \\
 & 5.6723 & 0.00279 & $-11.18 \pm 1.00$ \\
 & 5.7043 & 0.00276 & $-11.28 \pm 1.03$ \\
C & 6.3763 & 0.00221 & $-12.75 \pm 1.08$ \\
 & 6.4083 & 0.00219 & $-12.68 \pm 0.79$ \\
 & 6.4403 & 0.00217 & $-9.23 \pm 1.22$ \\
 & 6.5683 & 0.00208 & $-12.86 \pm 1.01$ \\
 & 6.6643 & 0.00202 & $-13.60 \pm 0.94$ \\
 & 6.7283 & 0.00199 & $-13.82 \pm 1.08$ \\
 & 6.7923 & 0.00195 & $-10.54 \pm 1.20$ \\
D & 10.2163 & 0.00086 & $-10.56 \pm 2.23$ \\
 & 10.2483 & 0.00086 & $-12.71 \pm 2.27$ \\
 & 10.2803 & 0.00085 & $-2.76 \pm 1.71$ \\
 & 10.4083 & 0.00083 & $-19.43 \pm 2.22$ \\
 & 10.5043 & 0.00081 & $-8.00 \pm 2.24$ \\
 & 10.5683 & 0.00080 & $-24.74 \pm 2.12$ \\
 & 10.6323 & 0.00080 & $-17.52 \pm 1.50$ \\
 & 10.6643 & 0.00079 & $-12.99 \pm 4.57$ \\
D (avgd) & 10.4563 & 0.00082 & $-13.59 \pm 2.49$ \\
\bottomrule
\end{tabular}
\end{table}

\end{appendix}

\end{document}